\documentclass{vietnam}

\def\Journal#1#2#3#4{{#1} {\bf #2}, #3 (#4)}

\def\apj{{\em ApJ}}
\def\aap{{\em A\&A}}
\def\mnras{{\em MNRAS}}

\def\be{\begin{equation}}
\def\ee{\end{equation}}
\def\bea{\begin{eqnarray}}
\def\eea{\end{eqnarray}}

\begin{document}
\title{Chemistry of Carbon-Chain Molecules in Star Forming Regions\\- Formation Pathway of HC$_{3}$N in G28.28-0.36 Hot Core -}

\author{Kotomi Taniguchi \& Masao Saito}

\address{Dept. of Astronomical Science, The Graduate University for Advanced Studies\\
Nobeyama Radio Observatory, National Astronomical Observatory of Japan, Nagano, Japan}

\maketitle\abstracts{
We carried out observations of three $^{13}$C isotopologues of HC$_{3}$N with the $J=9-8$ and $10-9$ rotational lines toward a hot core G28.28-0.36 with the 45-m radio telescope of the Nobeyama Radio Observatory in order to investigate the main formation pathway of HC$_{3}$N.
The abundance ratios are found to be 1.0 ($\pm 0.2$) : 1.00 : 1.47 ($\pm0.17$) ($1\sigma$) for $[$H$^{13}$CCCN$]: [$HC$^{13}$CCN$]: [$HCC$^{13}$CN].
From the results, we propose that the main formation pathway of HC$_{3}$N is the neutral-neutral reaction between C$_{2}$H$_{2}$ and CN.
We also compare the results among the different star-forming regions, from a low-mass starless core to a high-mass star-forming core.}

\section{Introduction}

\subsection{Carbon-Chain Molecules in Low-Mass Star-Forming Regions}

Carbon-chain molecules give us various information about star formation processes.
Carbon-chain molecules can be used as good indicators of starless and star-forming cores \cite{su,hi}.
They are abundant in young dark clouds, because they are efficiently formed by the gas-phase ion-molecule reactions.
In star-forming cores, they are deficient, while saturated complex organic molecules (COMs) are abundant, namely hot corino chemistry, which is similar to hot core chemistry for high-mass star-forming regions.

In contrast to hot corinos, Sakai {\it et al.}  detected various carbon-chain molecules in the low-mass star-forming region L1527 \cite{sa8}.
They proposed that CH$_{4}$ evaporated from grain mantles forms carbon-chain species efficiently.
Such chemistry was named warm carbon chain chemistry (WCCC). 
The short prestellar core phase would lead WCCC sources \cite{sa8}.
In this way, two scenarios of low-mass star formation, short and long prestellar core phases, have been proposed from the differences in the chemical composition in star-forming cores.

In molecular scale, the main formation pathways have been investigated at the cyanopolyyne peak in Taurus Molecular Cloud-1.
Deriving $^{13}$C isotopic fractionation, differences in abundances among the $^{13}$C isotopologues, is useful for tracing chemical reactions.
The main formation pathways of HC$_{3}$N \cite{ta} and HC$_{5}$N \cite{koa} have been revealed.
The main formation mechanism of HC$_{5}$N is the ion-molecule reactions between hydrocarbon ions (C$_{5}$H$_{n}^{+}$; $n=3-5$) and nitrogen atoms followed by electron recombination reactions \cite{koa}, whereas the neural-neutral reaction between C$_{2}$H$_{2}$ and CN was proposed as the main formation pathway of HC$_{3}$N \cite{ta}.    
We can directly study dominant chemical reactions from the observations deriving $^{13}$C isotopic fractionation.

Thus, the chemistry of carbon-chain molecules has been studied well from molecular level to chemical evolution in low-mass star-forming regions, and we obtain knowledge about the relationships between chemistry and physics during star formation processes.

\subsection{Carbon-Chain Molecules in High-Mass Star-Forming Regions}

While there are a lot of studies about carbon-chain molecules in low-mass star-forming regions, there were a few studies in high-mass star-forming regions.
Only in recent years, Green {\it et al.} detected HC$_{5}$N with the $J=12-11$ rotational lines (31.951777 GHz, $E_{\rm u}=10$ K) toward 35 hot cores with the Tidbinbilla 34-m telescope \cite{gr}. 
Their results seemed to support the chemical model calculation conducted by Chapman {\it et al.} previously \cite{ch}.
Chapman {\it et al.} showed that cyanopolyynes (HC$_{2n+1}$N; $n=1-4$) can be efficiently formed by C$_{2}$H$_{2}$ evaporated from grain mantles in hot core regions \cite{ch}.
However, the beam size of the Tidbinbilla 34-m telescope at the 32 GHz band was 0.95 arcmin, the $J=12-11$ rotational line can be easily excited in the dark clouds, and the excitation temperatures of HC$_{5}$N were not derived.
Their observational results then were not enough to confirm that HC$_{5}$N is formed in hot cores.
The chemistry of carbon-chain molecules has been hardly understood yet. 

\section{Our Research Aims}

Our studies aim to reveal the chemistry of carbon-chain molecules from molecular level to chemical evolution, applying the methods established in low-mass star-forming regions.
First, we are carrying out observations of carbon-chain molecules toward 17 high-mass starless cores (HMSCs) and 35 high-mass protostellar objects (HMPOs) with the Nobeyama 45-m radio telescope.
The main purpose of these survey observations is to investigate whether the $N$[HC$_{3}$N and/or HC$_{5}$N]/$N$[N$_{2}$H$^{+}$] ratios can be used as chemical evolutional tracers in high-mass star-forming regions.
Second, we carry out observations of long carbon-chain molecules toward hot cores with the Robert C. Byrd Green Bank Telescope (GBT) (K. Taniguchi et al., {\it {submitted to ApJ}}), the Karl G. Jansky Very Large Array (VLA) (K. Taniguchi et al., {\it {in prep.}}), and the Nobeyama 45-m radio telescope (K. Taniguchi et al., {\it {submitted to ApJ}}).
We have been doing data reductions and discussing now to establish new carbon-chain chemistry in hot core regions from these observations.
Third, the main formation pathway of HC$_{3}$N has been investigated toward the hot core G28.28-0.36.
In this proceeding, we mainly summarize this topic. 

\section{Investigation of Main Formation Pathway of HC$_{3}$N in a Hot Core G28.28-0.36}

We carried out observations of the three $^{13}$C isotopologues of HC$_{3}$N (H$^{13}$CCCN, HC$^{13}$CCN, and HCC$^{13}$CN) with the $J=9-8$ and $10-9$ rotational lines toward G28.28-0.36 with the Nobeyama 45-m telescope.
G28.28-0.36 is classified as the {\it Spitzer GLIMPSE} extended green objects (EGOs) \cite{cy}, and one of the hot cores where Green {\it et al.} detected HC$_{5}$N \cite{gr}.
Figure 1 shows the spectra of the three $^{13}$C isotopologues of HC$_{3}$N, and its normal species observed simultaneously \cite{kob}.
The abundance ratios are derived to be 1.0 ($\pm 0.2$) : 1.00 : 1.47 ($\pm0.17$) ($1\sigma$) for $[$H$^{13}$CCCN$]: [$HC$^{13}$CCN$]: [$HCC$^{13}$CN] \cite{kob}.
Therefore, the abundances of H$^{13}$CCCN and HC$^{13}$CCN are comparable with each other, and HCC$^{13}$CN is more abundant than the others.

\begin{figure}[ht]
\centering
$\begin{array}{cc}
\includegraphics[angle=0,height=8.cm]{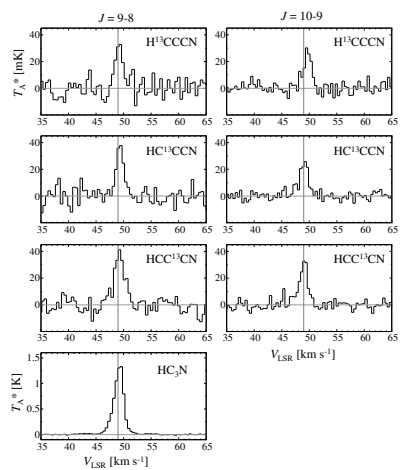} 
\end{array}$
\caption{\it {The spectra of the normal species and the three $^{13}$C isotopologues of HC$_{3}$N in G28.28-0.36 with the Nobeyama 45-m telescope.}}
\end{figure}

From the results, we discuss the main formation pathway of HC$_{3}$N in the hot core G28.28-0.36.
We investigate the possible formation pathways leading to HC$_{3}$N using the UMIST Database for Astrochemistry 2012 \cite{mc},  as shown in Figure 2.
There are four possible formation pathways as following;\\
\noindent Pathway 1: the ion-molecule reactions between C$_{3}$H$_{n}^{+}$ ($n=3-5$) and nitrogen atoms followed by electron recombination reactions,\\
\noindent Pathway 2: the ion-molecule reaction between HCN and C$_{2}$H$_{2}^{+}$  followed by electron recombination reaction,\\
\noindent Pathway 3: the neutral-neutral reaction between C$_{2}$H and HNC, and\\
\noindent Pathway 4: the neutral-neutral reaction between C$_{2}$H$_{2}$ and CN.\\
In Figure 1, we define $a:b:c$ for $[$H$^{13}$CCCN$]: [$HC$^{13}$CCN$]: [$HCC$^{13}$CN].
We also show the expected relationships among $a$, $b$, and $c$ by each formation pathway in Figure 2.
We find that  our observational results can match with only Pathway 4 (C$_{2}$H$_{2}$ + CN).
Therefore, the main formation pathway of HC$_{3}$N in G28.28-0.36 is the neutral-neutral reaction between C$_{2}$H$_{2}$ and CN.
This conclusion is consistent with the chemical model calculation \cite{ch}.
\\
\\
\begin{figure}[ht]
\centering
$\begin{array}{cc}
\includegraphics[angle=0,height=3.5cm]{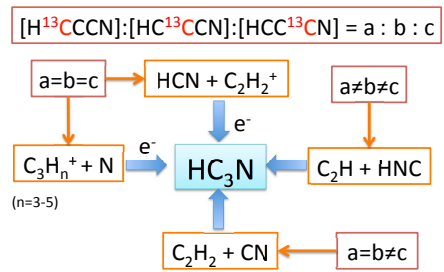} 
\end{array}$
\caption{\it {Possible Formation Pathways of HC$_{3}$N.}}
\end{figure}

\section{Comparison with Other Star-Forming Regions}

We summarize the observational results of $^{13}$C isotopic fractionation of HC$_{3}$N in various star-forming regions in Table \ref{tab:t1}.
We also carried out the same observations toward L1527 in order to compare between low-mass and high-mass star-forming cores \cite{kob}.
L1527 is one of the WCCC sources, and the main formation pathway of HC$_{3}$N was predicted by the chemical model calculation \cite{ha}.
Recently, other observations deriving $^{13}$C isotopic fractionation of HC$_{3}$N in L1527 were carried out \cite{ar}, and their results are consistent with our results within 1-sigma errors.
The results in TMC-1 \cite{ta} and Serpens South 1A \cite{li} are summarized as low-mass and high-mass starless cores, respectively.
We also show the typical temperature and density in each region in Table \ref{tab:t1}.

\begin{table}[htpb]
\caption{Comparison of $^{13}$C Isotopic Fractionation among Various Star-Forming Regions} \label{tab:t1}
\begin{tabular}{ccccccc}
\hline
 &  &  &  & source  & Temperature & Density\\
Source & H$^{13}$CCCN & HC$^{13}$CCN &HCC$^{13}$CN & type & [K] & [cm$^{-3}$] \\
\hline
L1527 \cite{kob} & 0.9 ($\pm 0.2$) & 1.00 & 1.29 ($\pm 0.19$) & WCCC & $20-30$ & $10^{5}$\\
G28.28-0.36 \cite{kob} & 1.0 ($\pm0.2$) & 1.00 & 1.47 ($\pm0.17$) & hot core & $100-200$ & $10^{6}$\\
TMC-1 CP \cite{ta} & 1.0 & 1.0 & 1.4 ($\pm0.2$) & dark cloud & 10 & $10^{4}$\\
Serpens South 1A \cite{li} & 0.91 ($\pm 0.09$) & 1.00 & 1.32 ($\pm 0.09$) & IRDC & 15 & $10^{5}$\\
\hline
\end{tabular}
\end{table}

We recognize the same $^{13}$C isotopic fractionation pattern in all of the four sources;\\
\noindent Characteristic 1: the abundances of H$^{13}$CCCN and HC$^{13}$CCN are comparable with each other, and\\
\noindent Characteristic 2: HCC$^{13}$CN is more abundant than the other two isotopologues.\\
As discussed in Section 3, this $^{13}$C isotopic fractionation pattern can be explained by the neutral-neutral reaction between C$_{2}$H$_{2}$ and CN.
In summary, the primary formation pathway of HC$_{3}$N may be common from low-mass prestellar cores to high-mass star-forming cores.

\section{Conclusions \& Future Works}

We carried out observations of the three $^{13}$C isotopologues of HC$_{3}$N in order to investigate its primary formation pathway in the hot core G28.28-0.36.
This study is the first one to apply the $^{13}$C isotopic fractionation method in both low-mass and high-mass star-forming cores.
The results suggest that the neutral-neutral reaction between C$_{2}$H$_{2}$ and CN is more dominant than the other formation pathways in both low-mass and high-mass star-forming cores, as well as starless cores.

As future works of the $^{13}$C isotopic fractionation method, we propose investigations about dependences on the spatial distributions.
We can study dependence of chemical reactions on the physical conditions in detail using the interferometry.
This method requires very high sensitivities, because we need to obtain spectra of $^{13}$C isotopologues with high signal-to-noise ratios.
ALMA will be able to solve the matter.

We also develop the studies about the chemistry of carbon-chain molecules established in low-mass star-forming regions into high-mass star-forming regions.
In particular, we can reveal the chemistry of carbon-chain molecules in hot cores soon.
This will provide us a new aspect of chemistry in hot core regions, which is not well understood today.
From these studies, we will investigate similarities and differences of the chemistry of carbon-chain molecules between high-mass and low-mass star-forming regions, in order to obtain information about a massive-star formation scenario.

\section*{Acknowledgments}
We would like to express our thanks to the staff of the Nobeyama Radio Observatory.
K. T. is most grateful to Dr. Tetsuhiro Minamidani, Dr. Yusuke Miyamoto, Dr. Hiroyuki Kaneko (NRO), Dr. Atsushi Nishimura (Nagoya Univ.), Dr. Tomoya Hirota, Dr. Fumitaka Nakamura (NAOJ), Dr. Kazuhito Dobashi, and Dr. Tomomi Shimoikura (Tokyo Gakugei Univ.) for fruitful advice and comments.
K. T. deeply appreciates Dr. Hiroyuki Ozeki (Toho University) discussing, advising and encouraging me.

\end{document}